\documentclass[a4paper,twocolumn,pra]{revtex4-1}
\usepackage{amsmath}
\usepackage{amssymb}
\usepackage{graphicx}
\usepackage{amsthm}
\usepackage{hyperref}

\begin{document}

\newcommand*{\cl}[1]{{\mathcal{#1}}}
\newcommand*{\bb}[1]{{\mathbb{#1}}}
\newcommand{\ket}[1]{|#1\rangle}
\newcommand{\bra}[1]{\langle#1|}
\newcommand{\inn}[2]{\langle#1|#2\rangle}
\newcommand{\proj}[2]{| #1 \rangle\!\langle #2 |}
\newcommand*{\tn}[1]{{\textnormal{#1}}}
\newcommand*{\1}{{\mathbb{1}}}
\newcommand{\T}{\mbox{$\textnormal{Tr}$}}
\newcommand{\todo}[1]{\textcolor[rgb]{0.99,0.1,0.3}{#1}}

\theoremstyle{plain}
\newtheorem{prop}{Proposition}
\newtheorem{proposition}{Proposition}
\newtheorem{theorem}{Theorem}
\newtheorem{lemma}[theorem]{Lemma}
\newtheorem{remark}{Remark}

\theoremstyle{definition}
\newtheorem{definition}{Definition}

\title{Relative Entropy as a Measure of Difference Between Hermitian and Non-Hermitian Systems}
\author{Kabgyun Jeong}
\email{kgjeong6@snu.ac.kr}
\affiliation{Research Institute of Mathematics, Seoul National University, Seoul 08826, Korea}
\affiliation{School of Computational Sciences, Korea Institute for Advanced Study, Seoul 02455, Korea}
\author{Kyu-Won Park}
\email{parkkw7777@gmail.com}
\affiliation{Department of Physics and Astronomy, Seoul National University, Seoul 08826, Korea}
\author{Jaewan Kim}
\email{jaewan@kias.re.kr}
\affiliation{School of Computational Sciences, Korea Institute for Advanced Study, Seoul 02455, Korea}

\date{\today}

\begin{abstract}
We employ the relative entropy as a measure to quantify the difference of eigenmodes between Hermitian and non-Hermitian systems in elliptic optical microcavities. We have found that the average value of the relative entropy in the range of the collective Lamb shift is large, while that in the range of self-energy is small. Furthermore, the weak and strong interactions in the non-Hermitian system exhibit rather different behaviors in term of the relative entropy, and thus it displays an obvious exchange of eigenmodes in the elliptic microcavity.
\end{abstract}

\maketitle

\section{Introduction}
The Hermitian property of observables is one of the fundamental principles in quantum physics. All the eigenvalues of a given Hermitian system are real and its eigenmodes corresponding to separable eigenvalues are orthogonal to each other. An ordinary closed system, such as a quantum billiard, has a Hermitian Hamiltonian with real eigenvalues~\cite{KKS99}. However, a real physical system cannot be completely isolated from its bath (or environment); it always interacts with the bath. Hence, the openness effect needs to be considered to study realistic physical models. A convenient and effective method to deal with an open system is the application of a non-Hermitian Hamiltonian model~\cite{D00,R09} in terms of the Feshbach projection-operator (FPO) formalism~\cite{F58}, in which the FPO divides the total Hermitian system into two subspaces of a non-Hermitian subsystem and a complementary subsystem known as bath.

The Hamiltonian operator of a non-Hermitian system typically has complex eigenvalues, and its eigenmodes corresponding to different eigenvalues are bi-orthogonal~\cite{L09} contrary to those in the Hermitian cases. Non-Hermitian Hamiltonians were originated from nuclear physics~\cite{F58} in 1958; nowadays they have been applied to diverse quantum mechanical systems not only in atomic~\cite{M11} and solid state physics~\cite{CK09}, but also for optical microcavities~\cite{PMJ+18,PMS+18}. Furthermore, non-Hermitian systems exhibit various physical phenomena such as phase rigidity~\cite{BRS07}, spontaneous emissions~\cite{S89,Lee+00}, parity--time symmetry~\cite{EMK+18,MPS15,FEG17}, exceptional points~\cite{XMJH16,FLE+20,COZ+17,MA19}, and Lamb shifts~\cite{PKJ16,PKJ+16}.

The Lamb shift describes the small energy difference in a quantum system due to system--bath coupling or vacuum fluctuations~\cite{LR47,SS10}. The effect was first studied in the case of the hydrogen atom~\cite{LR47}, and recently it has been investigated in metamaterial waveguides~\cite{YVR+09}, open photonic systems~\cite{LBDS18}, and optical microcavities~\cite{PKJ16,PKJ+16}.
In our previous works~\cite{PKJ16,PKJ+16}, we have employed the Lamb shift as a tool to systemically compare the Hermitian and non-Hermitian systems, by quantifying the difference between the energy eigenvalues of the Hermitian and non-Hermitian systems. However, certain appearance of the Lamb shift and the collective Lamb shift between two systems is very weak when the openness effects mostly appear in the imaginary part of the energy eigenvalues. Accordingly, we need to overcome this problem considering a different disparity, not of eigenvalues but of the eigenmodes in open systems. For this, we exploit the notion of relative entropy~\cite{KL51,NS20}, which is typically used to measure a difference given in the form of two probability distribution functions, to quantify the difference between the Hermitian and non-Hermitian eigenmodes. We could expect that adapting the relative entropy can be advantageous for future works on optical microcavity as in the cases of information theory.

In this article, we study such transitions for two-dimensional optical microcavities, which, owing to the isomorphic property of wave equations between optics and quantum physics, can be a good platform to investigate the wave-mechanical profiles~\cite{RBM+12,DMB+16}. While our system is a semi-classical, a fully quantum approach including quantum jump for the non-Hermitian system was recently introduced~\cite{MMCN19,AMMN20}.
The optical microcavities are very attractive resources for optoelectronic circuits both experimentally and theoretically~\cite{NS97,GCN+98}. They also present various physical properties such as a scar~\cite{HFD+03}, unidirectional emissions~\cite{RLK09,JZW+16,WH06}, high-quality factors~\cite{JXZ+12,SFL+09}, dynamical tunnelings~\cite{SGLX15,YLM+10,BKL+09}, ray dynamic~\cite{HS15}, and Lamb shifts~\cite{PKJ16,PKJ+16}. Specifically, we employed an elliptic microcavity as our open system, where it is appropriate for considering openness effects as the closed elliptic system is an integrable system and it clearly cannot result in any avoided crossings~\cite{S00}, as the avoided crossings are only induced by openness effects~\cite{PMJ+18,PMS+18}. Before presenting the main result, we briefly review the Hamiltonian of a non-Hermitian system and the two types of interactions in the optical microcavity.

\subsection{Non-Hermitian Hamiltonian and Two Types of Interactions}
This subsection provides basic notions of the non-Hermitian Hamiltonian of optical microcavities that are used in this article, and also discusses imaginary eigenvalues. Let us first consider a Hermitian Hamiltonian in a closed system $S$, known as a Hermitian system. As mentioned above, it always has real eigenvalues and its eigenmodes with different eigenvalues are orthogonal. Without the complete isolation of the system, it is natural to assume that the system interacts with its bath $B$; thus, the resulting system is open. In this case, the open system can be effectively described by a non-Hermitian Hamiltonian~\cite{D00,R09}.

The non-Hermitian Hamiltonian describing an open system can be defined by introducing FPOs $\Pi_S$ and $\Pi_B$ acting on the system subspace and on the bath, respectively, under the conditions~\cite{F58} $\Pi_S\Pi_B=0$ and $\Pi_S+\Pi_B=I_W$. Here, $\Pi_S$ is a projection onto the closed system whereas $\Pi_B$ is a projection onto its bath, and $I_W$ denotes an identity operator acting on the total space. With those projection operators and the Hermitian Hamiltonian for the total space (namely, $H_W$), we can easily define suitable operators such that $H_S=\Pi_SH_W\Pi_S$, $H_B=\Pi_BH_W\Pi_B$, $V_{BS}=\Pi_BH_W\Pi_S$, and $V_{SB}=\Pi_SH_W\Pi_B$. Here, $H_S$ and $H_B$ are the Hamiltonians of the closed system and the bath, respectively, and $V_{BS}$ and $V_{SB}$ denote the interaction Hamiltonians between the system and bath. Finally, we can formulate an \emph{effective} non-Hermitian Hamiltonian for this open system as~\cite{D00,R09}

\begin{align}
H_{\rm NH}=H_S+V_{SB}G_B^{(*)}V_{BS},
\label{eq1}
\end{align}
with an out-going Green function $G_B^{(*)}$ belonging to the bath. In general, the Hamiltonian $H_{\rm NH}$ has complex eigenvalues ($\zeta_j$), and its eigenmodes $\ket{\psi_{\rm NH}}$ are bi-orthogonal to each other. The bi-orthogonality condition can be represented by $\langle{\psi_j^L}|{\psi_k^R}\rangle=\delta_{jk}$ where $|{\psi_j^L}\rangle$ and $|{\psi_k^R}\rangle$ are left- and right-eigenmode, respectively~\cite{L09,B13}. It should be noted that the system Hamiltonian $H_S$ has real eigenvalues ($\lambda_j$) with orthogonal eigenmodes $\ket{\psi_j}_S$. In general, the non-Hermitian Hamiltonian in Eq.~(\ref{eq1}) has a multi-mode~\cite{D00,R09}, however we restrict that our system has only two-mode: that is, we consider interactions between two particular modes in which other modes are negligible. To study the openness effects, the matrix components of $H_{\rm NH}$ can be explicitly written separately with respect to the eigenbasis in $H_S$ as
\begin{align}
H_{\rm NH}=
\begin{pmatrix}
\eta_{11} & \eta_{12} \\
\eta_{21} & \eta_{22}
\end{pmatrix}+
\begin{pmatrix}
\delta_{11}  & \delta_{12}  \\
\delta_{21} & \delta_{22}
\end{pmatrix},
\end{align}
where the $\eta$- and $\delta$-components indicate the Hermitian part with real and the non-Hermitian parts with complex eigenvalues, respectively.

It is known that in the case of an integrable system $S$, the Hamiltonian $H_S$ has only diagonal $\eta_{jj}$ terms. Furthermore, to simplify the consideration of strong and weak interactions we also assume that the effective (non-Hermitian) Hamiltonian $H_{\rm NH}$ is symmetric and the off-diagonal (i.e., coupling) terms are real. Then, $H_{\rm NH}$ can be rewritten in matrix representation as
\begin{align}
H_{\rm NH}=
\begin{pmatrix}
\eta_{11}+\delta_{11} & \delta' \\
\delta' & \eta_{22}+\delta_{22}
\end{pmatrix},
\end{align}
where $\eta_{jj}$, $\delta'\in\mathbb{R}$, and $\delta_{jj}\in\mathbb{C}$. It should be noted that the diagonal terms ($\delta_{jj}$) correspond to the individual interactions of the energy levels with the bath; however, the off-diagonal or coupling terms ($\delta'$) correspond to the interaction of energy levels through the bath.
Thus, the eigenvalues of the effective Hamiltonian are given by
\begin{align}
\zeta_{\pm}=\frac{\omega_{1}+\omega_{2}}{2}\pm d,
\end{align}
where $\omega_{j}=\eta_{jj}+\delta_{jj}$ and $d=\sqrt{\frac{(\omega_{1}-\omega_{2})^2}{4}+\delta'^{2}}$. In the case of \emph{strong} interaction, i.e., $2\delta'>|\mathrm{Im}(\omega_{1})-\mathrm{Im}(\omega_{2})|$, the eigenvalues of $H_{\rm NH}$ show a repulsion in the real part (of complex energies) with a crossing at their imaginary part. Nevertheless, the eigenvalues of $H_{\rm NH}$ show a repulsion in the imaginary part while a crossing occurs in the real part in the case of $2\delta'<|\mathrm{Im}(\omega_{1})-\mathrm{Im}(\omega_{2})|$, i.e., the \emph{weak} interaction~\cite{W06}.

\begin{figure*}
\includegraphics[width=1.8\columnwidth]{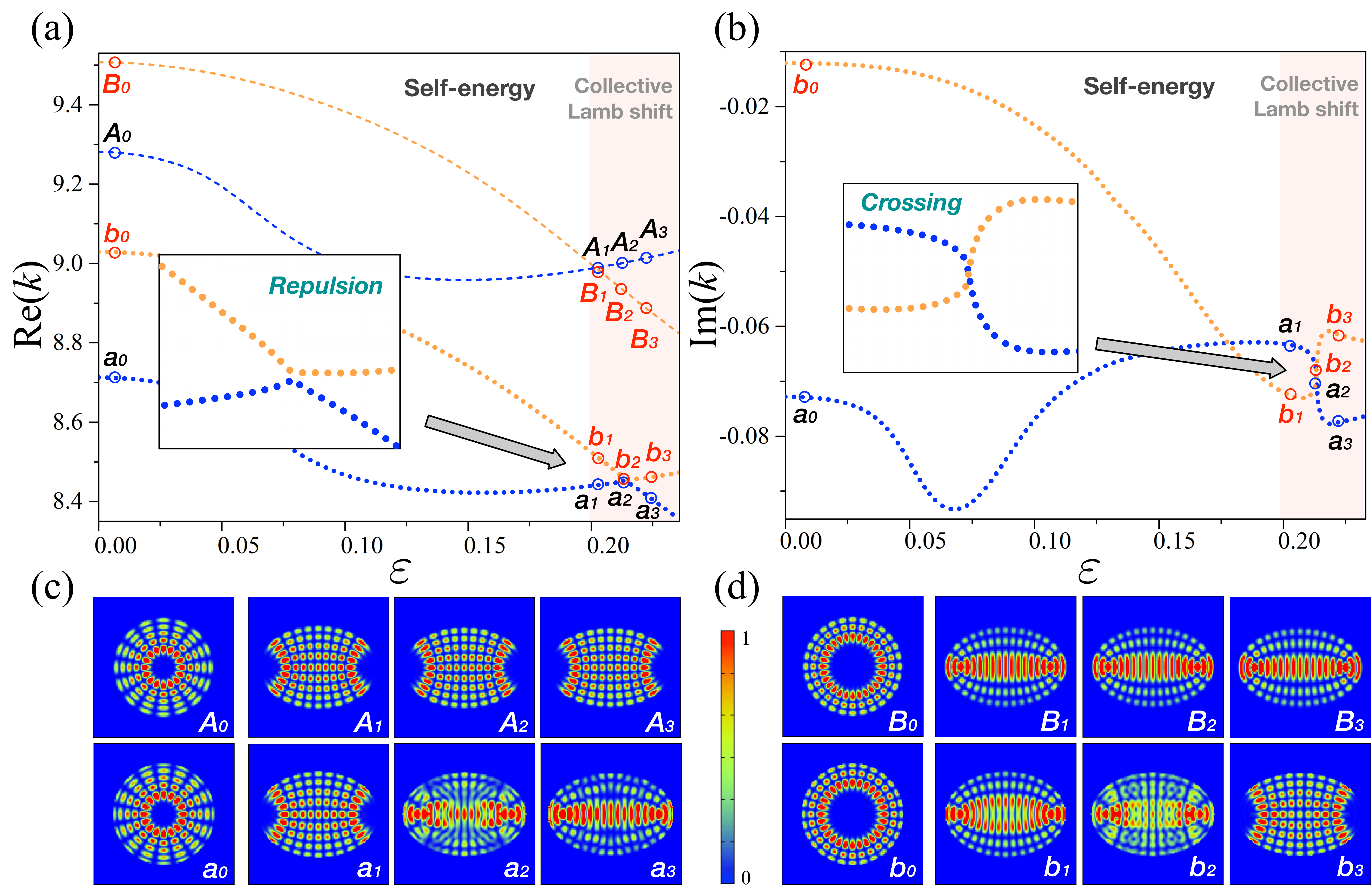}
\caption{Eigenvalues and their eigenmodes for strong interaction: the blue and orange lines correspond to level-$1$ and level-$2$, respectively. The thin dashed lines represent the closed system whereas the thick dotted lines indicate the open system. (a) Real part of the eigenvalues as a function of the deformation parameter $\varepsilon$. The thin dashed lines show a crossing while the thick dotted ones indicate a repulsion at $\varepsilon\approx 0.21$. (b) Imaginary part of the eigenvalues as a function of the deformation parameter $\varepsilon$. They show a crossing at $\varepsilon\approx 0.21$. (c) Configurations indicating mode intensities of certain representative eigenmodes for level-$1$. The shapes $a_{1}$ and $a_{3}$ indicate the exchange of the mode patterns. (d) Mode intensities of certain representative eigenmodes for level-$2$. They show similar trends to those in the plots in panel (c).}
\label{Fig1}
\end{figure*}

\section{Eigenvalues and Eigenmodes in Elliptic Optical Microcavities}
In the present work, we consider an elliptic optical microcavity as a model for the open system. The closed variant of this optical cavity belongs to an integrable system, and thus it cannot show any avoided crossings (i.e., repulsions)~\cite{S00}. In this framework, we can clearly observe openness effects related to the avoided crossings.  Here, we can obtain the eigenvalues and eigenmodes of the non-Hermitian Hamiltonian $H_{\rm NH}$ mentioned above only by solving the Helmholtz equation. Furthermore, we can plot the values by exploiting the boundary element method~\cite{W03} for transverse magnetic (TM) modes; we choose different boundary conditions satisfying $\psi(\textbf{r}=\Gamma)=0$ for the closed system, while $\psi_\tn{in}(\textbf{r}=\Gamma)=\psi_\tn{out}(\textbf{r}=\Gamma)$ and $\partial_{n}\psi_\tn{in}(\textbf{r}=\Gamma)=\partial_{n}\psi_\tn{out}(\textbf{r}=\Gamma)$ for the open system. Here, $\textbf{r}$ denotes a two-dimensional position vector, $\Gamma$ indicates a boundary of the optical cavity, and $\partial_{n}$ is the normal derivative. In our cases, the Helmholtz equation is naturally given by $\nabla^{2}\psi+n^{2}k^{2}\psi=0$, where $\psi$ denotes the vertical component of the electric field, $n$ is the refractive index of the optical cavity, and $k$ is the wave number. It should be noted that resonances ($\psi$) and their wave numbers ($k$) in the Helmholtz equation play the role of eigenmodes and their eigenvalues, respectively.

In the following, we carefully analyze certain eigenvalue trajectories and eigenmode patterns for the effective non-Hermitian Hamiltonian $N_{\rm NH}$ under two categories, that is, the strong and weak interactions in the elliptic optical microcavity. In this section, first we discuss the strong interaction.

\subsection{Strong Interaction}
For a given effective Hamiltonian $H_{\rm NH}$, if any repulsion is observed between the energy levels in the real part (not in imaginary) regime, the interaction is usually considered to be strong.
Figure~\ref{Fig1} shows a number of representative eigenvalue trajectories and eigenmode patterns of the effective Hamiltonian for the strong interaction under the deformation parameter $\varepsilon\in[0,0.23]$ with refractive index at $n=2.825$ (fixed) in the elliptic optical microcavity. The parameter $\varepsilon$ is directly connected to the major axis ($a=1+\varepsilon$) and the minor axis ($b=\frac{1}{1+\varepsilon}$) of those ellipses, and the area of all ellipses is maintained as $\pi$. For convenience, we consider a two-level system only. Here, we label the blue colors as ``level-$1$'' and the orange colors as ``level-$2$'' in both cases for the open and closed systems. The eigenmodes for level-$1$ at $\varepsilon=0$ (circle) has a radial quantum number $l=5$ and an angular quantum number $m=8$, beside the eigenmodes for level-$2$ at $\varepsilon=0$ (circle) has radial quantum number $l=3$ and angular quantum number $m=14$, respectively. Furthermore, the thin dashed lines represent the eigenvalues of the closed system whereas the thick dotted lines represent those of the open system. Figure~\ref{Fig1}(a) shows the real part of the eigenvalues of both the closed and open systems as a function of the deformation parameter $\varepsilon$. The thin dashed lines (i.e., closed system) exhibit a crossing, while the thick dotted (i.e., open system) exhibit a repulsion at $\varepsilon\approx 0.215$. There is no repulsion of eigenvalues in the closed elliptic system as it belongs to an integrable system, which has no interaction in the system. Thus, this clearly confirms that the repulsion of eigenvalues in an open elliptic system completely results from the effects of openness. The imaginary part of the eigenvalues of the open system is also plotted in Figure~\ref{Fig1}(b), where they exhibit a crossing at $\varepsilon\approx 0.215$. This indicates the essential feature for manifesting strong interaction, as mentioned above.

In addition, Figures~\ref{Fig1}(c) and~\ref{Fig1}(d) show explicit eigenmode patterns for level-$1$ and level-$2$, respectively. The subtle change in the eigenmode patterns in the closed system is comparable to those in the open system at the narrow region, known as \emph{collective Lamb shift}, i.e., $\varepsilon\in[0.2,0.23]$. This change is due to the variation of the geometric boundary conditions. Nevertheless, an open system shows a rather different aspect. That is, the eigenmodes corresponding to $a_{1}$ and $b_{1}$ are definitely mixed at the center of the avoided crossing (i.e., $a_{2}$ and $b_{2}$) and then they are exchanged at $a_{3}$ and $b_{3}$. This exchange of the eigenmodes across the avoided crossing is also a necessary feature for the strong interaction. In the following section we discuss the weak interaction for the effective non-Hermitian Hamiltonian in an open system.

\begin{figure*}
\includegraphics[width=1.8\columnwidth]{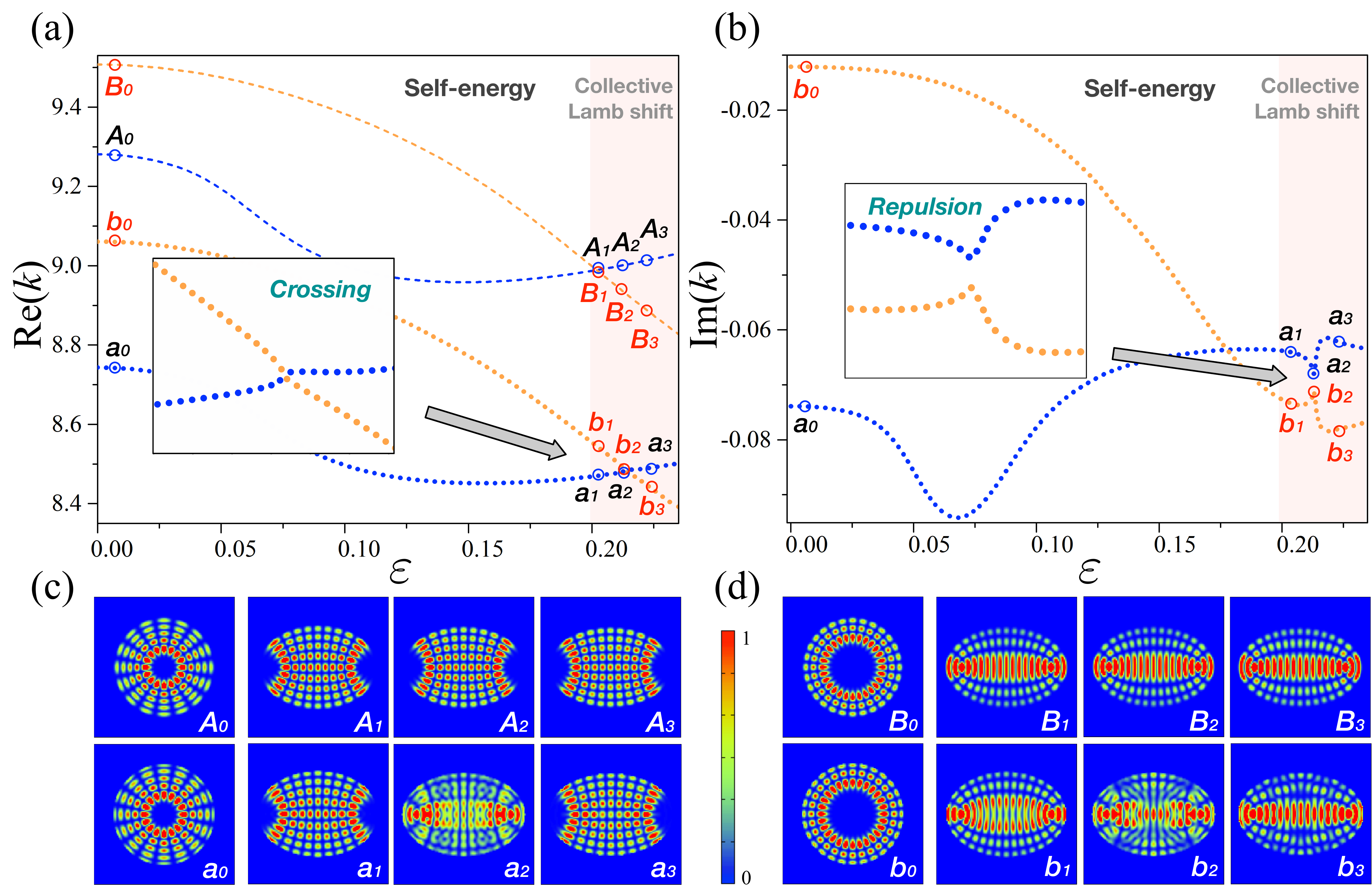}
\caption{Eigenvalues and their eigenmodes for weak interaction. (a) Real part of eigenvalues as a function of the deformation parameter $\varepsilon$. The thick dotted lines show a crossing. (b) Imaginary part of eigenvalues as a function of the deformation parameter $\varepsilon$. They show a repulsion at $\varepsilon\approx 0.21$. (c) Configurations denoting mode intensities of certain representative eigenmodes for level-$1$. The shapes $a_{1}$ and $a_{3}$, indicate the non-exchange of the mode patterns. (d) Mode intensities of certain representative eigenmodes for level-$2$. They show similar trends to those of the plots in panel (c).}
\label{Fig2}
\end{figure*}

\subsection{Weak Interaction}
As a complementary part of the strong interaction, Figure~\ref{Fig2} shows the weak interaction in terms of eigenvalue trajectories and eigenmode patterns with $\varepsilon\in[0,0.23]$ and the constant refractive index $n=2.82$. Apparently, Figure~\ref{Fig2} similar to Figure~\ref{Fig1}. However, directly opposite properties can be recognized from the magnified view of the insets of Figures~\ref{Fig1} and~\ref{Fig2}. The thick dotted lines in Figure~\ref{Fig2}(a) clearly reveal a crossing at $\varepsilon\approx 0.215$, while those in Figure~\ref{Fig2}(b) show a repulsion at $\varepsilon\approx 0.215$. Actually, these phenomena are the main characteristics of the eigenvalues to define the weak interaction itself. Moreover, the overall morphologies of the mode pair ($a_{1}$,$a_{3}$) in Figure~\ref{Fig2}(c) are similar to each other. The mode pair ($b_{1}$,$b_{3}$) in Figure~\ref{Fig2}(d) exposes similar trends to those in Figure~\ref{Fig2}(c). These properties indicate that there are not any exchanges of the eigenmodes in the weak interaction regime.

In addition, we can presume that an exceptional point (EP) is located in the range of $n\in[2.82, 2.825]$, as the EP is a singular point in the parameter space in which a transition between the weak and strong interactions takes place. Indeed, the EP is located near at $n\approx 2.8238$ and $\varepsilon\approx 0.2134$ (not shown). It should also be noted that the two-eigenmode patterns (i.e., $a_{2}$ and $b_{2}$) are similar to each other as they lie near to the EP.

\begin{figure}
\includegraphics[width=1.0\columnwidth]{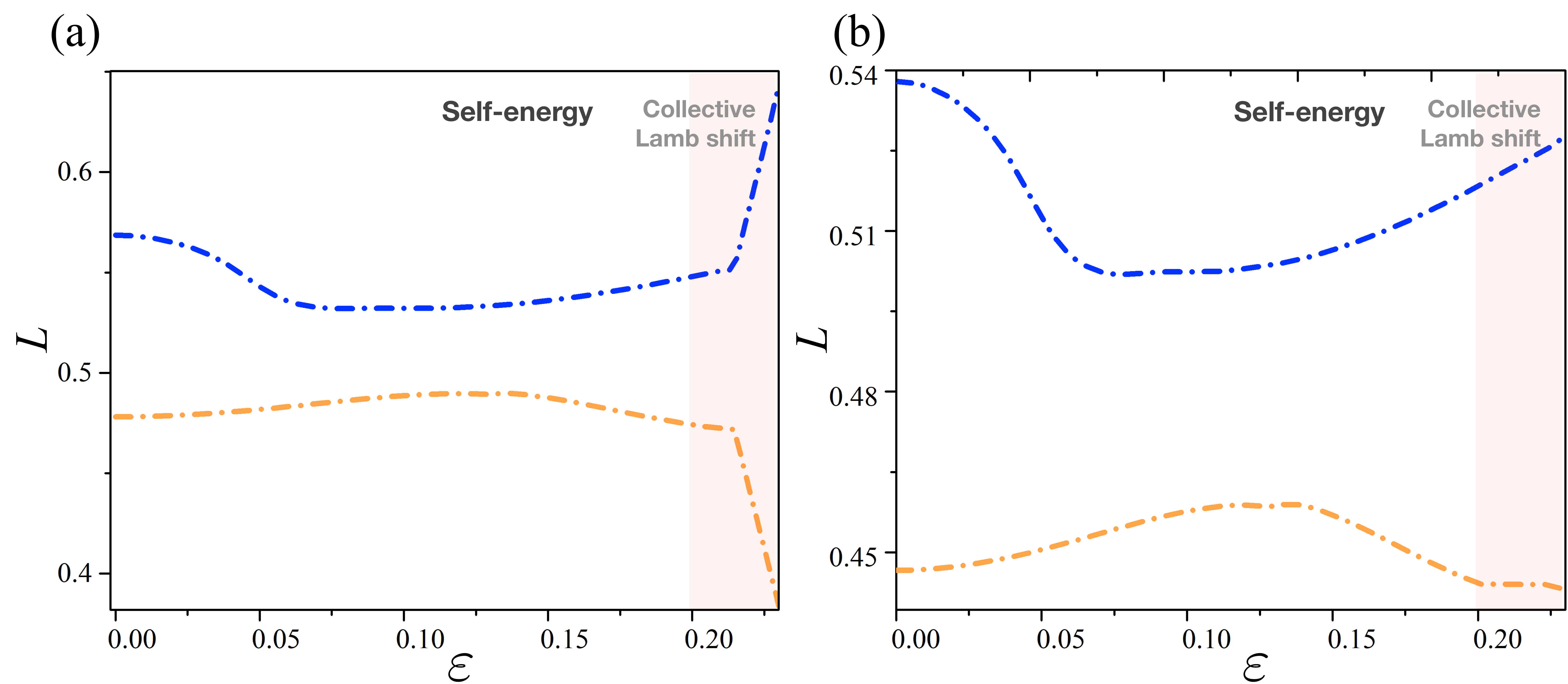}
\caption {Lamb shifts in the strong and weak interactions as a function of the deformation parameter $\varepsilon$. (a) Lamb shift $L$ in the strong interaction in which the shift is changed drastically across the avoided crossing. (b) Lamb shift $L$ in the weak interaction changed slightly across the avoided crossing.}
\label{Fig3}
\end{figure}

\section{Lamb Shifts in Elliptic Optical Microcavity}
In this section, we discuss the Lamb shift, which is a small difference of energy levels in the elliptic optical microcavities.
As it has been found recently, there are two kinds of Lamb shifts~\cite{SS10,R13}. The first type is known as the \emph{self-energy} or simply Lamb shift in atomic physics: an energy-level shift occurs due to the individual interaction of energy levels with their bath. The second type is the \emph{collective Lamb shift}: it is an energy-level shift arising from the interaction of energy levels with each other via the external bath~\cite{R13}.

In the framework of non-Hermitian systems, the Lamb shift is defined by the differences between the real eigenvalues ($\lambda_{j}$) of $H_{\rm S}$ and the real part of complex eigenvalues ($\zeta_{j}$) of $H_{\rm NH}$. For any $j$, it is defined by
\begin{align}
L_{j}= \lambda_{j}-{\rm Re}(\zeta_{j}).
\end{align}
Considering the matrix elements of the non-Hermitian Hamiltonian, it should be noted that the diagonal components of the non-Hermitian Hamiltonian ($\delta_{jj}$) result in the self-energy (i.e., detuning), while the off-diagonal components ($\delta_{jk}$) of the Hamiltonian matrix result in the collective Lamb shift (i.e., avoided crossing)~\cite{R13,PMJ+18}.

As shown in Figure~\ref{Fig3}, when the deformation parameter $\varepsilon$ is varied, the Lamb shift $L$ in the strong and weak interaction regions can be plotted as a function of $\varepsilon$. The Lamb shift (dashed-dot line) shown in Figure~\ref{Fig3} is just the difference between the same colored curves in Figures~\ref{Fig1} and~\ref{Fig2}, i.e., the dashed orange (blue) lines and dotted orange (blue) lines. Figure~\ref{Fig3}(a) shows the Lamb shift in the strong interaction region. In that case, $L$ is changed drastically across the avoided crossing. On the contrary, the Lamb shift in the weak interaction region, shown in Figure~\ref{Fig3}(b), only changed slightly, compared to that shown in Figure~\ref{Fig3}(a). This is because the avoided crossing takes place at the real part in the strong interaction range, while the avoided crossing occurs at the imaginary part in the weak interaction range. Although the Lamb shift formally describes the openness effects of a system, the phenomenon cannot capture the essential feature of the weak interactions easily, as it only deals with the energy-level shift; the Hermitian Hamiltonian does not have any imaginary part of the energy comparable to the non-Hermitian Hamiltonian. Therefore, we need to consider a general method to quantify the difference between the Hermitian and non-Hermitian systems using an entirely different measure for entropic quantities.

\begin{figure*}
\includegraphics[width=1.9\columnwidth]{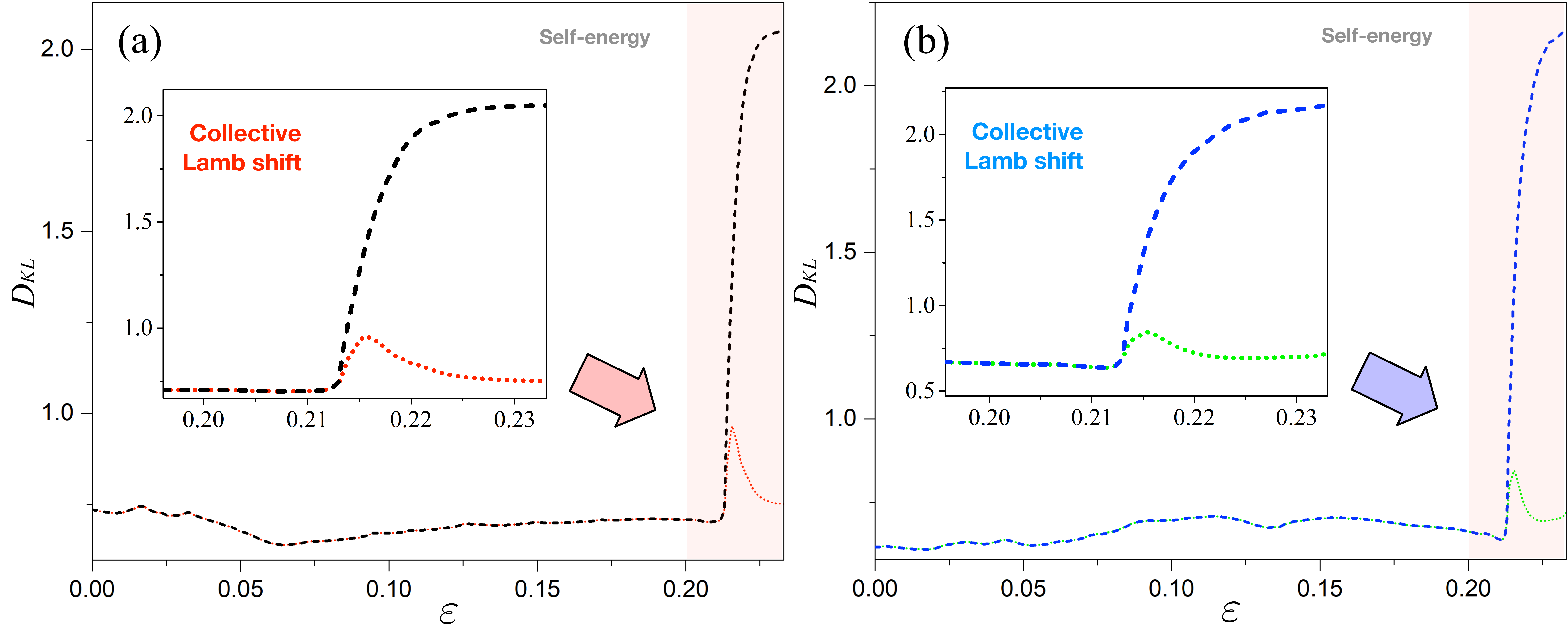}
\caption {Relative entropy of eigenmodes: the relative entropy $D_{\rm KL}$ is plotted as a function of the deformation parameter $\varepsilon$. (a) For level-1, the black dashed lines indicate the strong interaction and the red dotted lines represent the weak interaction. The relative entropy in the self-energy range has lower average values compared to that in the collective Lamb shift. (b) For level-2 it has similar performance to that in level-1.}
\label{Fig4}
\end{figure*}

\section{Relative Entropy for Hermitian and Non-Hermitian Systems}

As discussed above, Lamb shifts only on the eigenvalues have a limit to describe the nature of openness. Here, we propose that comparing eigenmodes is a more generic and appropriate method for describing a difference between Hermitian and non-Hermitian systems. For this reason, we introduce the notion of relative entropy (also known as the Kullback--Leibler divergence)~\cite{KL51,NS20}. The relative entropy is a measure of the difference in two probability distributions on the probability space. For any discrete probability distributions $P$ and $Q$, the relative entropy is generally defined as
\begin{align}
D_{\rm KL}(P\| Q)= \sum_{j=1}^{N}P(x_j)\log \frac{P(x_j)}{Q(x_j)},
\end{align}
where $N$ is the total number of elements in the discrete probability distribution.

First, to apply the relative entropy to the optical system, the probability distributions on $Q(x_j)$ and $P(x_j)$ need to be defined for any $j$. These probability distributions can be obtained by considering the normalized mode patterns corresponding to the each eigenmodes in the Hermitian and non-Hermitian systems, respectively; that is, $\sum_{j=1}^{N}|\psi_{S}(x_j)|^{2}=\sum_{j=1}^{N}Q(x_j)=1$ and $\sum_{j=1}^{N}|\psi_{\rm NH}(x_j)|^{2}=\sum_{j=1}^{N}P(x_j)=1$. Here, $N$ is a mesh point of the discretization of the area within the cavity into $N$ pieces. With these normalized mode patterns (i.e., probability distributions), we can numerically calculate the relative entropies of each eigenmodes in the cavity; the results are shown in Figure~\ref{Fig4}.

\subsection{Relative Entropy in Self-energy}
Figure~\ref{Fig4} shows the relative entropies $D_{\rm KL}$ for the eigenmodes as a function of the deformation parameter $\varepsilon$. For level-1, the black dashed lines represent calculations under strong interaction and the red dotted lines represent results under weak interaction while for level-2 the blue dashed lines and the green dotted lines represent results from strong and weak interactions, respectively. The relative entropy exhibits rather different behaviors according to the different openness effects: both of the two $D_{\rm KL}$ entropies in the self-energy range (i.e., $\varepsilon\in[0,0.2]$) have lower average values compared to that in collective Lamb shift. They are moving around the value of $D_{\rm KL}\approx 0.5$ and varied slightly depending on the parameter $\varepsilon$. These results are not unexpected, as it can be clearly seen that the overall morphologies of two eigenmode patterns $(a_{0}$ and $A_{0}$) are similar to each other. Likewise, the pairs of eigenmode patterns of ($a_{1}$,$A_{1}$), ($b_{0}$,$B_{0}$), and ($b_{1}$,$B_{1}$) are individually similar to each other. This pattern occurs due to the fact that the self-energy in an open system can be regarded as an interaction with the bath itself, and it only changes the boundary condition of the eigenmode on the system, resulting in a decay mode with nearly invariant overall morphology.

\subsection{Relative Entropy in Collective Lamb Shift}
Contrary to the relative entropies in the self-energy region, those in the collective Lamb shift region exhibit diverse behaviors as shown in the insets of Figure~\ref{Fig4}. In the case of weak interaction, both dotted lines (red and green) are maximized at the center of the avoided crossing ($\varepsilon\approx 0.215$) with the value of $D_{\rm KL}\approx 0.8$. In other words, it can be verified that the two values of the relative entropy at the ends of the avoided crossing do not change significantly, which can be explained as follows. The interactions of eigenmodes via the bath result in, without modes exchanging, a coherent superposition of eigenmodes as well as a mixing of them. As a result, the mixed eigenmodes $a_{2}$ and $b_{2}$ in the non-Hermitian system are considerably changed compared to the eigenmodes $A_{2}$ and $B_{2}$ in the Hermitian system, and then the non-exchange of the eigenmodes ($a_{3}$ or $b_{3}$) across the avoided crossing reduces the relative entropy.

Nevertheless, the curves of the relative entropies under strong interaction (dashed lines) keep increasing, passing the center of the avoided crossing and they are saturated to larger values rather than those far before the avoided crossing. This behavior clearly reveals the exchange of eigenmodes from $a_{1}$ ($b_{1}$) to $a_{3}$ ($b_{3}$) shown in Figure~\ref{Fig1}; thus, we can quantitatively describe the exchange of eigenmodes. It should also be noted that at the center of the avoided crossing, the value of $D_{\rm KL}$ in the strong interaction (dashed lines) is lager than that in the weak interaction (dotted lines). This confirms that the strong interaction is actually stronger than the weak one. Thus, the difference between the Hermitian and non-Hermitian systems can be quantified by exploiting the relative entropy, even when the Lamb shift cannot provide any significant information on it.

\begin{figure*}
\includegraphics[width=2\columnwidth]{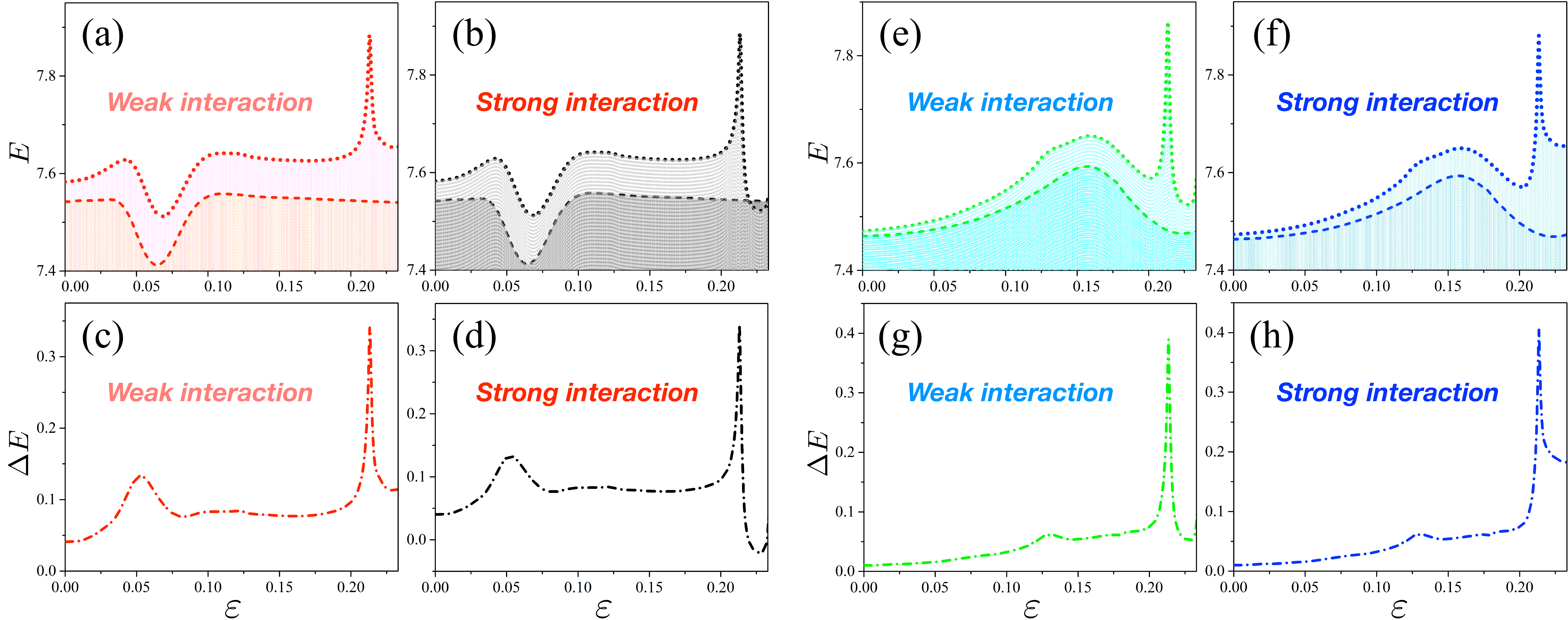}
\caption {Comparative differences of Shannon entropies for Hermitian and non-Hermitian systems. (a) Shannon entropies of level-$1$ for open (red dots) and closed (red dashed) systems in the weak interaction. (b) Shannon entropies of level-$1$ for open (black dot) and closed (black dash) system in the strong interaction. (c) Comparative difference of two kinds of Shannon entropies in panel (a). (d) Comparative difference between two kinds of Shannon entropies in panel (b). (e) Shannon entropy of level-$2$ for open (green dots) and closed (green dashed) systems under weak interaction. (f) Shannon entropy of level-$2$ for open (blue dot) and closed (blue dash) systems under strong interaction. (g) Comparative difference between two kinds of Shannon entropies in panel (e). (h) Comparative difference between two kinds of Shannon entropies in panel (b).}
\label{Fig5}
\end{figure*}

\section{Comparative Difference of Shannon Entropies}
In general, the relative entropy is not equal to the comparative difference of Shannon entropies $E$, and they are equal to each other only when the probability $Q$ has a uniform distribution~\cite{KL51}, i.e., $Q=\frac{1}{N}$. This fact can be confirmed in our non-Hermitian system. Each dotted lines in panels (a), (b), (e), and (f) of Figure~\ref{Fig5} denotes the Shannon entropy of $H_{\rm NH}$ for the non-Hermitian system, and it is defined by $E_{\rm NH}(P)=-\sum_{j=1}^NP(x_j)\log P(x_j)$. Each dashed lines represents $E_S(Q)=-\sum_{j=1}^NQ(x_j)\log Q(x_j)$ for its corresponding Hermitian system, and each dot-dashed lines in panels (c), (d), (g), and (f) in Figure~\ref{Fig5} corresponds to their comparative difference, $\Delta E(P:Q)=E_{\rm NH}(P)-E_S(Q)$.

As can be seen clearly, $\Delta E$ exhibits very different properties with respect to the relative entropy $D_{KL}$. For example, the average of the $\Delta E$ values is less than that of the $D_{\rm KL}$ values, and at the center of the avoided crossing, $\Delta E$ in the strong interaction is similar to that in the weak interaction, contrary to the relative entropy.

\section{Discussion}
We studied the relative entropy to quantify the difference of eigenmodes between the Hermitian (closed) and non-Hermitian (open) systems in an elliptic optical microcavity. Although, our previous work employing Lamb shifts studied this question, the result cannot be applied directly to the weak interaction regime. In this study, we improved the limit by investigating the interactions not on the eigenvalues but on the eigenmodes under the same optical settings. 

Especially, the average value of the relative entropy in the collective Lamb shift is larger than that of in the self-energy region. In addition, at the center of the avoided crossing the value of the relative entropy in the strong interaction is larger than that in the weak interaction regime. This directly confirms that the strong interaction is actually stronger than the weak interaction in the optical microcavity. We also studied the dramatic exchanges of eigenmodes quantitatively by employing the relative entropy. Specifically, the exchange of eigenmodes takes place when the relative entropy keeps increasing across the avoided crossing. In addition, we confirmed that a comparative difference of the Shannon entropies is different from the relative entropy in both cases of the strong and weak interactions.

There are still several interesting questions about the non-Hermitian systems in the optical microcavities. In future studies, quantum versions of entropy, such as von Neumann entropy and quantum mutual information, can be considered instead of the classical cases, such as Shannon or relative entropies, for quantum entanglement in the cavity. We expect that those kinds of researches can extend our current knowledge on quantum information and quantum sensing technologies beyond the studies on the optical aspects only.

\section*{Acknowledgments}
This work was supported by the National Research Foundation of Korea, a grant funded by the Ministry of Science and ICT (NRF-2020M3E4A1077861) and the Ministry of Education (NRF-2018R1D1A1B07047512). K.-W.P. acknowledges support from Samsung Science and Technology Foundation under Project No. SSTF-BA1502-05. J.K. was supported in part by KIAS Advanced Research Program CG014604. We thank the Korea Institute for Advanced Study for providing computing resources (KIAS Center for Advanced Computation Linux Cluster) for this work.


%

\end{document}